\documentclass[runningheads]{llncs}
\pdfoutput=1
\usepackage{graphicx}
\usepackage{subfig}
\usepackage{enumitem}
\usepackage[utf8]{inputenc}

\begin{document}
%
\title{Parsisanj: a semi-automatic component-based approach towards search engine evaluation.\thanks{Supported by Iran Telecommunication Research Center.}}
\titlerunning{Parsisanj: a component-based approach towards SE evaluation.}
%
\author{Amin Heydari Alashti\inst{1}\orcidID{0000-0002-3689-5796} \and
Ahmad Asgharian Rezaei\inst{2}\orcidID{0000-0002-8510-8875} \and
Alireza Elahi\inst{3} \and
Sobhan Sayyaran\inst{4} \and
Mohammad Ghodsi\inst{5}
}
\authorrunning{A. Heydari Alashti et al.}
%
\institute{
BigData Solutions Land, Iran \\
\email{heydari@rahkar.co}\and
RMIT University \\
\email{ahmad.asgharian.rezaei@rmit.edu.au}\and
Shahid Beheshti University, Tehran, Iran \and
Imam Sadegh University, Tehran, Iran \and
Computer Science Facutly, Sharif University of Technology, Tehran, Iran
}
\maketitle              

\begin{abstract}
Accessing to required data on the internet is wide via search engines in the last two decades owing to the huge amount of available data and the high rate of new data is generating daily. Accordingly, search engines are encouraged to make the most valuable existing data on the web searchable. Knowing how to handle a large amount of data in each step of a search engines' procedure from crawling to indexing and ranking is just one of the challenges that a professional search engine should solve. Moreover, it should also have the best practices in handling users' traffics, state-of-the-art natural language processing tools, and should also address many other challenges on the edge of science and technology. As a result, evaluating these systems is too challenging due to the level of internal complexity they have, and is crucial for finding the improvement path of the existing system. Therefore, an evaluation procedure is a normal subsystem of a search engine that has the role of building its roadmap. Recently, several countries have developed national search engine programs to build an infrastructure to provide special services based on their needs on the available data of their language on the web. This research is conducted accordingly to enlighten the advancement path of two Iranian national search engines: Yooz and Parsijoo in comparison with two international ones, Google and Bing. Unlike related work, it is a semi-automatic method to evaluate the search engines at the first pace. 
Eventually, we obtained some interesting results which based on them the component-based improvement roadmap of national search engines could be illustrated concretely. 
\end{abstract}
\section{Introduction}
Internet growth is meaningfully related to the information needs of users. The more information gets available on the internet is the direct result of the more information needs raises. Thus, these needs should be addressed in an appropriate way regarding the huge amount of data that should be processed. Additionally, there are a plethora of uncountable categories in which the data is produced. Therefore, some systems which could join the information requirements of users with the available data on the internet are formed. Those systems could achieve an acceptable level of natural language understanding as they go further. They are called search engines; a multi-component system with exactly defined responsibilities, for instance, web crawler, parser, indexer, ranker. Knowing how each of these components work is a key point to find how to enhance their performance and precision. 
Enlightening the roadmap of search engines' advancement is the main reason for their evaluation, and the other goals such as finding a suitable domain-specific search engine for a specific goal can be in a lower priority relatively. So, to truly address the main aim of search engine evaluation, it is crucial to analyze their components one by one in detail. Unlikely, previous work assumed a search engine as a black-box system; thus, their final results contained rank component evaluation and are not practically effective for improving search engines.  
Mostly, manually evaluating the ranking component, researchers are forced to choose a small set of queries to docile this method's high costs. Additionally, some semi-automatic work is strictly dependent on search engines' log, and also their evaluation range covers the ranking component for navigational queries. In the present work, a semi-automatic component-based search engine's evaluation method is proposed. According to our knowledge, it is a break-through contribution in comparison with all previous similar systems in terms of the number of queries, query types coverage, evaluation methods, low cost, result consistency and reliability. 

The remaining of this paper is organized as follows: related work is introduced and discussed in section 2, the whole structure of Parsisanj is illustrated in section 3. The evaluation domains and their sub-categories at each component are introduced in section 4. Evaluation Features, score functions, and the final scores of each search engine are discussed from sections 5 to 7 respectively. 

\section{Related Work}
Previous work can be categorized differently based on different features like the structure of the query-set, the query-set size, the evaluation type, and the study generality level. 
\subsection{Structure of Query-set}
The query-set structure is modified using various approaches like putting constrain to use specific query types, different construction methodologies, and their query sources. Some like \cite{mahmoudi_1} constrained their query-set to navigational queries to compare the performance of search engines. Some other studies that used a mixed set of informational and navigational queries like \cite{sanchez_5}. IR datasets like TREC are one of the query sources of which some researchers utilize for extracting their query-set.\cite{wu_7}  

But the most prevalent method of building query-set in previous work is selecting a set of keywords by crowd-sourcing and extracting keywords from available documents like academic papers' keywords, search engines' search log files. \cite{mahmoudi_1,azimzadeh_2,farzane_3,nowkarizi_4,sanchez_5,zhang_6,rahim_8,gul_9,t_10}

\subsection{Query-set Size}
Most of the studies use a small set of queries to control their manually evaluation system's costs. It turns out that the type of the query is an important factor that impacts the size of the query-set. Accordingly, the only set which has an acceptable size is \cite{mahmoudi_1} that contains 2000 queries, but all the queries are navigational. The next greatest query size is for \cite{farzane_3} which utilized crowd-sourcing and selected a subset of search engines' query log that contains 400 queries; however, usually search engine's query log is not an accessible source for anyone. \cite{wu_7} A query set size of 200 is the next which selected them from the TREC dataset. Other studies built a set of keywords by crowd-sourcing or extracting paper keywords to build their datasets, so they were unable to make a thorough evaluation because of their small query size. 
\cite{farzane_3,nowkarizi_4,zhang_6,wu_7,rahim_8,gul_9}

But as a matter of fact, due to the large amount of data gathered and processed in search engines, and the complex architecture of components it contains, it would be impossible to evaluate their behavior with just a small set of queries. Previous work has assessed a small portion of search engines' components which cannot give a complete illustration of weak and strong points of them. Thus, to achieve the goal of building a general roadmap for improving search engines it needs to use much more well-defined queries for evaluating search engines. 

\subsection{Automatic/Manual Evaluation}
This feature is again highly related to the query types of each study. \cite{mahmoudi_1} has provided a set of 2000 Persian navigational queries and submit them into Google, Bing, and Parsijoo. Although it had the largest query-set, due to the nature of navigational queries that have just a single correct answer, the assessment can be done automatically. Of course, the assessment's aim and available meta-data are the other evaluation type specifiers factors. For instance, \cite{nowkarizi_4} assessed the overlap and coverage of search engines, so it can test its studying search engines' results by aggregating and comparing the returning links of search engines automatically. 
On the other hand, \cite{sanchez_5} uses some computational linguistics' datasets to evaluate the hit count returned by search engines for each query. But it used a manual approach to test if its automatic evaluation method is reliable or not. 

\subsection{Study Generality level}
As mentioned before, due to the small set of queries the previous related studies had, they could only test a small portion of the domains which a search engine covers. They have just tested the ambiguity level in queries that search engines can handle \cite{wu_7}; the coverage of special-purpose websites, like national language websites and websites that are active in a special science category, by considering navigational queries \cite{mahmoudi_1,rahim_8}; finding a search engine that returns the most robust \textit{hit count} property for computational linguistics' research domains \cite{sanchez_5}. 

\section{Parsisanj Structure}
According to the related work section, there are some problems in design and assumptions in building the structure of their evaluation methods: 
\begin{itemize}
	\item[$\bullet$] using a small set of features cannot illustrate a real view of a large and multi-aspect system like a search engine. 
	\item[$\bullet$] assessing manually will bring subjectivity in evaluation results. 
	\item[$\bullet$] regardless of appraising a search engine according to user view-point, ranking is not the only component that should be assessed. Actually, there are also other components that will impact users' experience too. 
\end{itemize}

Accordingly, we tried to make a novel evaluation structure based on the components a search engine usually has. The workflow of designing such a system is as follow: 
\begin{enumerate}
	\item illustrating the components of a well-designed search engine in detail. 
	\item designing evaluation domains which are the structure of evaluating each component with different levels of difficulty. 
	\item designing metrics by which the relevancy score of each query with search engines' results is measured. 
	\item depicting the step by step roadmap by which the query-set should be designed.
\end{enumerate}
Each of the above steps is built based on the gained knowledge in its previous step. Generally, by finding a good picture of a search engine's architecture, we built a roadmap for assessing each component. A detailed designed query is submitted, then ambushes for special results specified in the query-set design step from the query. 
\subsection{Detected Components}
In this study, a search engine is divided into 
\subsubsection{Query Analyzer}
Each input of a search engine is transformed into a predefined form by preprocessing steps of the query analyzer. On the other hand, the input is a fetched webpage or users' queries that might have various forms that need to be unified. This unifying step prepares input data for further processes. The query analyzer component consists of sub-systems like text normalizer, tokenizer, spell correction, query expansion which are discussed in detail in the following part. 
\paragraph{Text normalizer} is responsible to add or remove some parts of the terms in the input text to modify it to a common form between all the input types of the search engine using a well-defined mapping function. In Persian text, some similar Arabic characters can be used interchangeably with Persian ones. Moreover, some Arabic characters are entered in the Persian language, and its side effects are making multiple written forms for a single word. Besides, about 70\% of Persian alphabets have similar shapes and sounds. As a result, the multi-shaped words with a single meaning or multiple meanings is an important issue in Persian text preprocessing.
\paragraph{Tokenizer} is responsible for splitting an input text to its meaningful finer granularities like paragraphs, sentences, phrases, and terms. In a search engine, finding these smaller parts is a key point to build an efficient and effective index and ranking process. Persian has eight basic structure of verbs and multiple types of compound verbs.\cite{givi_11} Additionally, it is among highly inflectional languages that can produce different kinds of verbs and syntactic phrases using its inflectional rules. Noun and adjective phrases have some special features in Persian text which makes them much more challenging. Ezafe forms noun phrases, but the problem is that it is not written and is just pronounced. Therefore, tokenizing Persian text is a challenging task that needs facing various challenges. 
\paragraph{Spell Correction} checks if there is any term which is out of language's vocabulary, or if a term does not match the context. Input text of search engines from both directions(user input, fetched webpages) may have typos. Having typos means increasing false negatives in the matching process. So, finding suitable solutions can bring us many more matches for users' queries. Furthermore, by finding a typo in an input, spell correction comes in action to suggest a list of terms for substituting it. The process of selecting terms in the list is too important for making an error-free system. 
\paragraph{Query expansion} is responsible for moving a general context query to a more specific context using various information additional information. Firstly, it was just based on some ontologies like wordnet and language models that were built on the web data. Secondly, it moves on using gathered information from users and helped search engines to propose a personalized search result. Adding some terms and phrases to a query to make it more specific is too risky which can result in increased false positive. \\
There are solutions for the above-mentioned query analyzer sub-systems challenges. Noticing the noisy nature of web data which is the main input of a search engine, achieving a suitable solution for building a great design and implementation of the above systems is much more challenging. 
\subsubsection{Ranking}
The rank component constructs the interface of a search engine by presenting search results of a query based on top of the output of the previous steps of the engines' pipeline. Although errors of previous steps can be spread to this step, we assume that previous steps are errorless in evaluation time. When a query is sent to a search engine, the query analyzer will process it and finds user intention represented in some terms and a feature set. 
The processed query is used to select a candidate list of pages to be the input of the rank component. Generally, the rank component uses a complex score function to reach an outperforming combination of the pages and the query features. All the previous work concentrated on evaluating this component manually which makes the results subjective. They had multiple problems in their assumptions: 1. ranking is not the only component of a search engine that can affect the final result, so others should be evaluated too. 2. other components are not directly in touch with the final ranked list, though they can be evaluated directly by a carefully designed query-set. 3. automatically evaluating the rank component of a search engine does not mean the implementation of its complex score function. 
\section{Evaluation Domains}
Evaluation domains define the type and specifications of queries for each of the components precisely. This is one of the main contributions of this work which covers most of the critical challenges the components face within experimental environments. Moreover, it has a hierarchy of difficulty which helps us to measure the level of expertise in each component of search engines. In the following part evaluation domains of each sub-system of 
components will be discussed: 
\subsection{Text Normalizer}
In general, this sub-system should map multi-shaped terms to a single shape to increase text-matching accuracy.
\begin{enumerate}
\item Mapping numbers to written form and vice-versa
	\begin{enumerate}
	\item[1.1] Cardinal numbers
	\item[1.2] Ordinal numbers
	\item[1.3] Cost and benefits
	\item[1.4] Time
	\item[1.5] Date
	\item[1.6] Population
	\end{enumerate}
\item Single words with multiple written forms
	\begin{enumerate}
	\item[2.1] Hamzeh based multi-form words
	\item[2.2] Character repetitions with similar sounds
	\item[2.3] Detecting correct character's initial, centric and final forms
	\end{enumerate}
\item Words with a single sound but different written forms
	\begin{enumerate}
	\item[3.1] All are live words
	\item[3.2] Just one form is live
	\end{enumerate}
\end{enumerate}

\subsection{Text Tokenizer}
This sub-system splits input text to the needed granularity level which like terms, phrases, sentences and paragraphs. 
\begin{enumerate}
\item terms are joint without seperator
\item phrase detection
	\begin{enumerate}
	\item[2.1] two part verbs
	\item[2.2] three to five part verbs that at least one of them has plural suffix
	\item[2.3] named entities prepended by identifiers
	\end{enumerate}
\end{enumerate}

\subsection{Spell Correction}
Spell correction should find typos in a query to increase the chance of correct matching in the index component. 
\begin{enumerate}
\item Lexicon
\item Inflection
\item Homonyms
\item Frequency of words
\item Keyboard order
\end{enumerate}

\subsection{Query Expansion}
Expands a query using the below items to build a more specific query. 
\begin{enumerate}
\item Synonyms
\item Abbreviations
\item Punctuations
\end{enumerate}

\subsection{Ranking}
Ranks fetched results from the index component based on the information needs of the user extracted from the query. 
\begin{enumerate}
\item Navigational Queries
\item Trends with single URL
\item Known items
\end{enumerate}


\section{Evaluation Features}
Features help us to build an automatic and highly precise query evaluation system. The fundamental webpage relevancy check based on a query is searching for the query's appointed features on its content. Combining these features makes the total structure of this system's score functions. Features of various types can be divided into the below categories: 
\begin{enumerate}
\item Content-based: occurrence of metrics and their frequency, content's length, ...
\item Structure-based: cares about the occurrence of content-based metrics in different parts of a webpage disparately. 
\item Based on result sets structure: inverse document frequency(IDF), and mean reciprocal ranking(MRR) are from this type which considers a result set's structure to evaluate a page's score. 
\item Webpage's domain-based: authority and hub domains are always more trustful than regular ones. 
\item Hybrid: a combination of the above categories can result in a general evaluation scenario for each component. Content-based and structure-based metrics can be divided into this type.
\end{enumerate}
Moreover, the features' value is calculated based on their type: 
\begin{enumerate}
\item If they have a concrete method of calculation, so they are \textit{shallow features}. E.g. publish time, the occurrence of a specific script in page content, age of the host, Alexa rank, ...
\item If they need some information from the \textit{universal set}\footnote{In our system, the universal set contains a set of related webpages to a query. They are used to elicit a value interval for query features by which relevancy of a webpage can be distinguished in terms of each of its features.}, then they are inference based features. The universal set\footnote{U set} members are selected based on their close relationship with a query. Thus, attributes and aspects of a relevant result can be extracted using it. But the \textit{decision network} provides the utility to find the best combination of these attributes besides applying different importance coefficient of each attribute in it. Some examples of these features are the occurrence of descriptive terms, the occurrence of exclusive terms, document length, URL depth, document readability and ...
\end{enumerate}

\section{Score Functions}
In this section the score function that is based on the decision network is discussed. The general form of our score function is in eq.\ref{eq1}: \\
\begin{equation}\label{eq1}
S_\Lambda(D;Q) = \Sigma_j\lambda_j . f_j(D,Q) + Z
\end{equation}
In the above equation, $f_j(D,Q)$ is feature j that evaluates document D according to query Q. $\lambda_j$ is the importance coefficient of feature j; and $\Lambda$ is the collection of all the $\lambda$s. $\Lambda$ is calculated using: 
\\
\begin{equation}\label{eq2}
argmax_\Lambda E(R_\Lambda;T)
\end{equation}
$E(R_\Lambda;T)$ is the assessment parameter between the score of results given by Parsisanj's score function($R_\Lambda$) and the score of Results given by an expert(T). It can be the amount of convergence of the score function to the assessment done by an expert. In other words, features' coefficients($\Lambda$) should be tuned somehow to converge the score function's output to expert assessments on a test-set. The size of the test-set is estimated using \textit{hypothesis testing} to ensure that the assessment parameter can represent an ideal set of $\Lambda$. 

\subsection{Hypothesis testing}
Various factors affect a feature's score; and the most challenging type of features are the hybrid ones. Thus, to handle these complexities, we used decision networks. A decision network is a directed acyclic weighted network that its nodes can represent hybrid metrics and their dependencies; so a node can represent the occurrence of a phrase in content or specifically in one of its sub-parts. It also should be mentioned that based on Markov chain theory, the list of descriptive and exclusive terms of each query is enriched by bigram combinations of their tokens. 

\begin{figure}
\includegraphics[width=\textwidth]{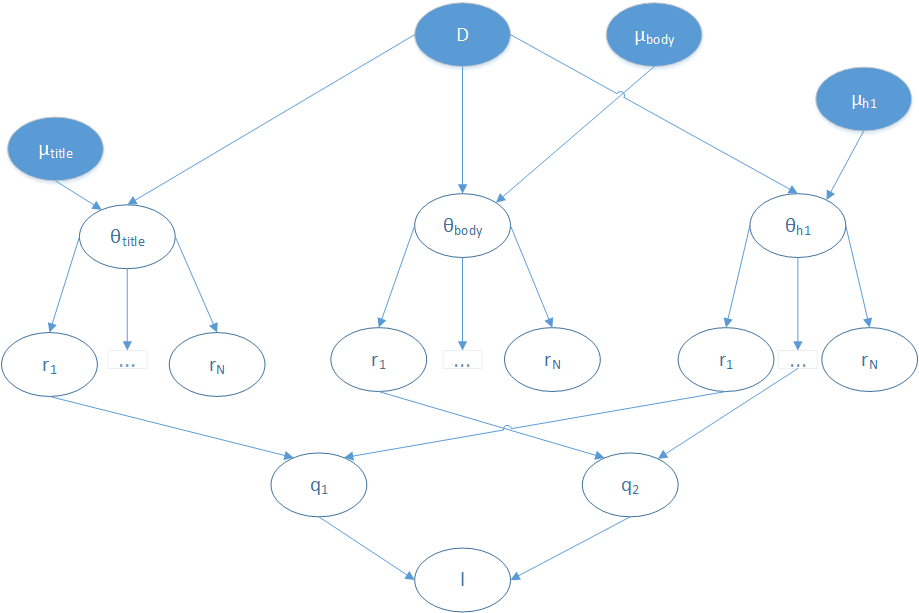}
\caption{Decision network of a document D that its relevancy to the queries $q_1$ and $q_2$ is evaluated using metrics $r_1$,$r_2$,...,$r_n$.}\label{fig1}
\end{figure}

In figure \ref{fig1}, $\Theta$s are external dependencies that are calculated according to related documents in the U set of a query. As an example, in searching a descriptive term in a document, if the number of its occurrences on related documents of the U set is between 10-15, then it is anticipated to observe such a similar behavior in other relevant documents too. As mentioned before, $\Theta$ is the distribution that is extracted from the U set.\\
U set is built at the query set creation time by experts. They add relevant pages to a query from search engines that are not under the evaluation. The root of the decision tree is the evaluating document; and its children are different parts of the document. Each part contains its related part of the U set. For instance, the title part of the document contains the title of the U set pages. Moreover, nodes of each part of the document($\Theta_{body}$) are connected to the features($r_i$) that can be defined in that part. 

The relevancy score calculation process for a page using a decision network is as follows: each feature($r_i$) in the network has a corresponding random variable($X_i$). The value of the random variable shows whether the feature has occurred in the page content or not. The following paragraphs discuss the method of setting the random variable's value. Furthermore, each feature has an importance coefficient($W_i$), and each part of a page has its special importance coefficient($V_{\theta_{i}}$). The final relevancy score between the query and the document is: 
\begin{equation}
DocRelevancyScore = \Sigma_{i,\theta_{i}\in\Theta} X_i.W_i.V_{\theta_{i}}
\end{equation}
The method for calculating the value of the random variable is straightforward based on its definition; and each part of the document's importance is the random variables' value that is normalized by summation of the random variable for the U set pages. E.g. the following formulas are for calculating the value and weight of the random variable correspondence to occurrence frequency of a descriptive term in title of a page: 
\begin{equation}
value=count(m;d)
\end{equation}
\begin{equation}
weight=count(m;d)/\Sigma_{u \in U} count(m;u)
\end{equation}

The closest part of Parsisanj to a real search engine is its score function. It might be a suspicion that Parsisanj tries to implement the score function of a search engine; hence, its ranking is not fair and reliable. But there are key differences between what Parsisanj does and what search engines do; which makes the mentioned assumption false. A list of differences is presented in Table \ref{tab2}. 

\begin{table}
\begin{center}
\caption{The key differences between Parsisanj's score function and search engines' ranking component.}\label{tab2}
\begin{tabular}{|p{3cm}|p{4.5cm}|p{4.5cm}|}
\hline
\textbf{title} &  \textbf{Parsisanj} & \textbf{Search Engine}\\
\hline
involving features &  predefined per query & extracting on the fly based on query's information need\\
\hline
\#processing pages &  a limited number of the top results of each query & walks
 through whole its index for each query\\
 \hline
\end{tabular}
\end{center}
\end{table}

Firstly, table \ref{tab2} shows that there is no need to be a search engine to have a great ranking algorithm; In other words, Parsisanj moved manually evaluation of search engines from evaluating the result pages of a set of query to the very first step of designing the query set. Therefore, it can evaluate search results of a large array of search engines without any further cost, and simultaneously diminishes the risk of subjectivity in evaluating search engines. \\
Secondly, the small amount of evaluating result pages helps to be much faster in the relevancy check process, and consequently, it can test various score functions to improve its evaluation process.

\section{Results}
Hereafter, we will discuss the evaluation outcomes that were invoked in two phases for each of the search engines; each of these two contains some more fine-grained steps in which we will illustrate weak and power points of search engines. 

Query Analyzer's modules evaluation part evaluates about 63 thousand result pages of queries. It includes evaluating Normalizer, Tokenizer, Query expansion, and SpellChecker modules of search engines. 
\subsection{Normalizer}
By and large, it was believed that the performance of most of the sub-tasks in the text normalization step has a direct relation to the amount of language-specific knowledge is utilized in designing the systems. 
\subsubsection{Conversion between numbers and their written form}
figure~\ref{fig2}:a shows that all the search engines have a fundamental problem in converting numbers to their written form and vice-versa. However, supporting the normalization step of all the languages is not expected from international search engines, a coverage of less than eight percent is too disappointing. Regardless of the international search engines' results, it is obvious that the two national ones have not cared about this step. Consequently, it can be the source of further performance missings in other downstream tasks like ranking. 

\subsubsection{Words with multiple written forms}
figure~\ref{fig2}:b shows that Google and Bing can find different written forms of words regardless of being a multi-lingual international search engine. Yooz and Parsijoo are close in this step, but they have not achieved Google's and Bing's performance. Moreover, it can confirm that utilizing a huge amount of data beside a robust statistical method can result in an excellent level of performance in this normalization subtask. 
\subsubsection{Homophones} 
\begin{figure}%
    \centering
    \subfloat[numbers and their written form]{{\includegraphics[width=6.2cm]{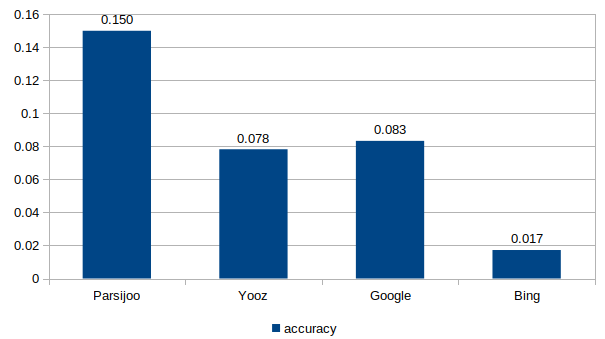} }}%
    \subfloat[words with multiple written forms]{{\includegraphics[width=6.2cm]{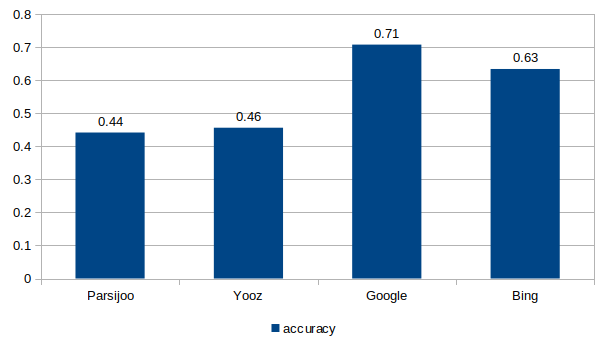} }}%
    \qquad
    \subfloat[homophones]{{\includegraphics[width=6.2cm]{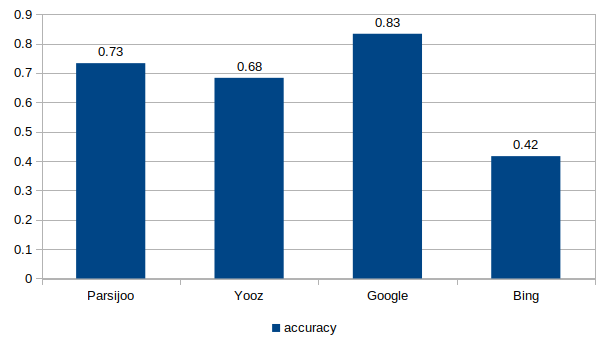}}}
    \subfloat[sum up]{{\includegraphics[width=6.2cm]{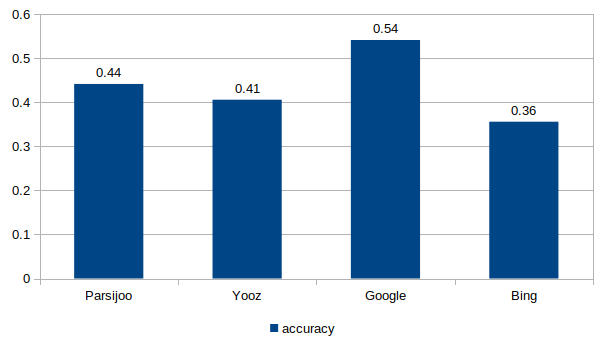}}}
    \caption{Normalizer evaluation}%
    \label{fig2}%
\end{figure}

Bing in comparison with the other three search engines, have a much lower level of expertise in handling Persian homophones(figure~\ref{fig2}:c). However, the national search engines results are much lower than Google's result which might have not put any language-specific knowledge in these modules. 

To sum up, in figure~\ref{fig2}:d it is obvious that Google has made a better normalization pipeline for Persian contents. The second best search engine is Parsijoo; by regarding its much lower index size, it made a great job in comparison with Google's normalization score. Additionally, with a mere difference in score Yooz is chasing Parsijoo. In comparison with the first three search engines, Bing could not achieve acceptable performance in the normalization pipeline. Eventually, Google's results reject our first assumption about needed language-specific background knowledge for addressing normalization tasks. 

\subsection{Tokenizer}
\begin{figure}%
    \centering
    \includegraphics[width=6.2cm]{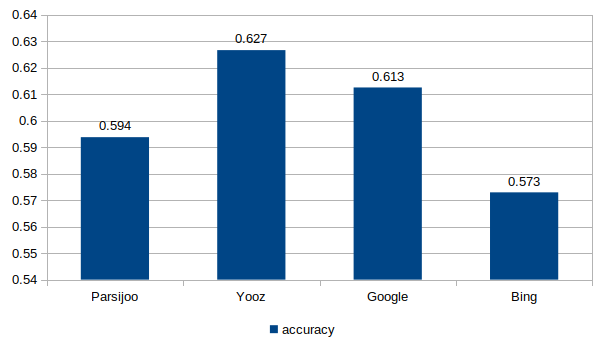}
    \caption{Tokenizer evaluation}%
    \label{fig3}%
\end{figure}

This part's queries consist of distinguishing concatenated words and detecting multi-word verbs. In figure~\ref{fig3}, the performance of the search engines is presented. Google as the best and Parsijoo as the second-best search engine solved this challenge. It is probably due to the utilization of a rich language model. Furthermore, none of the engines could reach an acceptable performance in detecting multi-word verbs, so the results of this type of questions are at the minimum in figure~\ref{fig3}.

\subsection{Spell Correction} 
\begin{figure}%
    \centering
    \includegraphics[width=6.2cm]{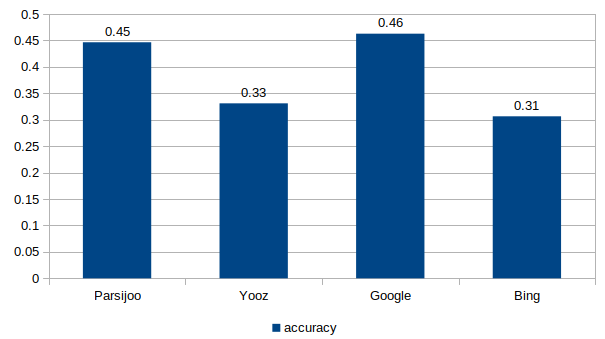}
    \caption{Spell Correction's evaluation}%
    \label{fig4}%
\end{figure}

figure~\ref{fig4} shows the accuracy of spell correction module of search engines. Parsijoo has designed a much more robust spell correction system rather than Yooz and Bing. It might have achieved a better result even better than Google if it had a huge amount of data Google Utilizes. Bing shows that besides not having language specific considerations for non-English spell correction, its design lacks the ability to detoured such tasks using statistical methods. 

\subsection{Query Expansion}
\subsubsection{Expanding query with synonyms set} In this task performance of a system can be easily influenced by the amount of indexed data in the search engine. So as it is expected, international search engines achieved better results than the two national ones. In figure~\ref{fig5}:a, Google and Bing are meaningfully better than the other two which can be a sign of having a much better and greater amount of data. Additionally, undertaking more sophisticated algorithms to expand a query is helps boost their results in this level. 
\subsubsection{Handling abbreviations} 
\begin{figure}%
    \centering
    \subfloat[using synonyms]{{\includegraphics[width=6.2cm]{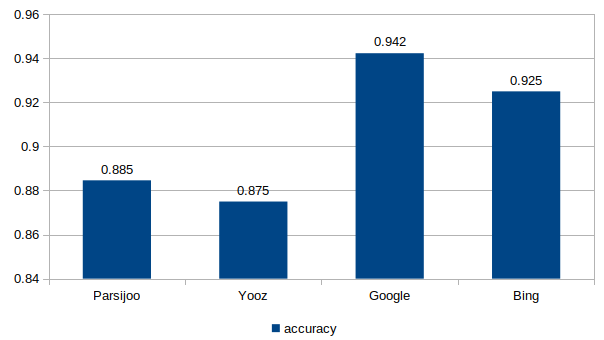} }}%
    \subfloat[handling abbreviations]{{\includegraphics[width=6.2cm]{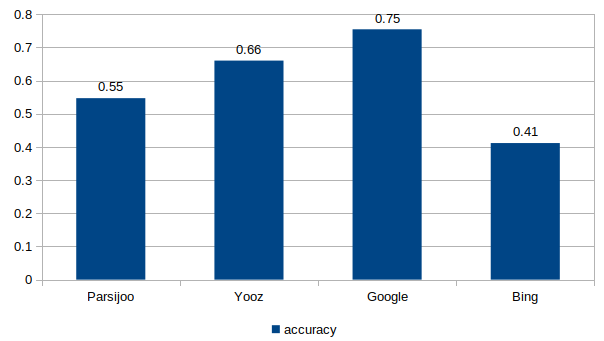} }}
    \qquad
    \subfloat[sum up]{{\includegraphics[width=6.2cm]{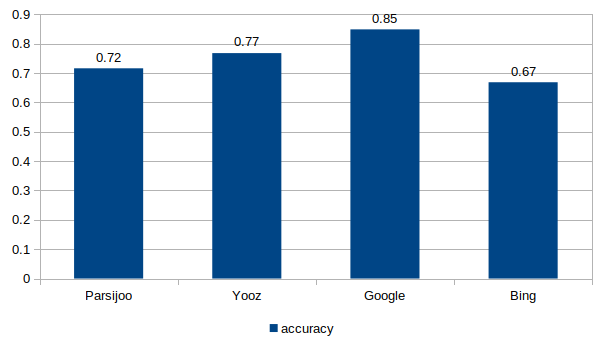}}}
    \caption{Query expansion evaluation}%
    \label{fig5}%
\end{figure}

Same as expanding a query using synonyms, the amount of indexed data plays an important role to distinguish abbreviations. Moreover, results show that access to a huge amount of data is crucial but not enough to design a robust system. Figure~\ref{fig5}:b, shows that Google and Yooz have a close and acceptable level of supporting abbreviations in a query. Bing's accuracy reveals that regardless of its access to a huge number of indexed pages, it may suffer from not having a sophisticated algorithm for detecting abbreviations. 

The overall overview given by figure~\ref{fig5}:c shows that in general handling synonyms is addressed more than abbreviations in search engines. Abbreviations vanished the score of expansion using synonyms for Bing. On the other side, two national search engines compensate their score by supporting abbreviations much better than Bing. 

\begin{figure}
\centering 
\includegraphics[width=6.2cm]{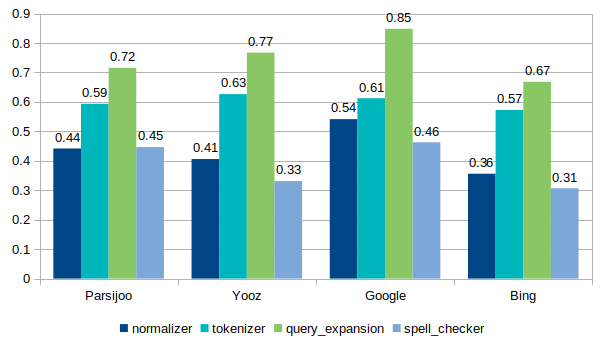}
\caption{Phase1 total scores}\label{fig6}
\end{figure}

\subsubsection{Query Analyzer} figure~\ref{fig6} shows that Google has the best overall query processor component among the evaluated search engines. It confirms that most of the fundamental search engine tasks can be addressed greatly using language-independent methods. Additionally, Yooz, Parsijoo are at a similar level at this phase with a mere difference in their total score of some tasks. Bing gained the lowest score in all the tasks, however, its scores are generally close to other search engines. 

\subsection{Rank}
In the second phase, about 100 thousand result pages of about 5 thousand queries were fetched and evaluated. The query types that are designed for evaluating this phase are navigational, known items, and semi-informational queries. Navigational is the type of query that the user knows the goal website domain and searches to find that exact one. Known items are similar to navigational, but the query is asking for knowledge not a website domain. Semi-informational queries may have multiple correct answers, e.g. searching for the recipe of food which may have some similar recipes. The decision network is used to evaluate the relevancy of the search engine's result pages for the last two types of queries. Different parts of a document are specified in the network and their scores are aggregated as described before in this document. 

\subsubsection{Navigational queries} score of search engines are illustrated in figure~\ref{fig7}:a. Google, Bing, and Parsijoo's result in covering this type of queries are similar to each other. However, Google has a giant crawler, Parsijoo has achieved an acceptable score. On the other side, Yooz is far away from the other search engines. It is too disappointing for a search engine; however, it can be the result of a problem in their web crawler or ranking algorithm. 

\subsubsection{Known items} the accuracy score of all the search engines are in an acceptable range, although Yooz is not as well as others. There is a great difference between the value of Mean Reciprocal Rank(MRR) and signed MRR\footnote{If an irrelevant page is ranked higher than the relevant results of a query the search engine will receive -1 score for that query.} bars with the accuracy score. It shows that relevant answers to this type of query are not among the top-ranked results of search engines. Thus, although search engines like Bing are doing their best in this type of query, they should improve their ranking algorithm to ameliorate their MRR score. 

\subsubsection{Semi-informational} 
\begin{figure}%
    \centering
    \subfloat[navigational]{{\includegraphics[width=6.2cm]{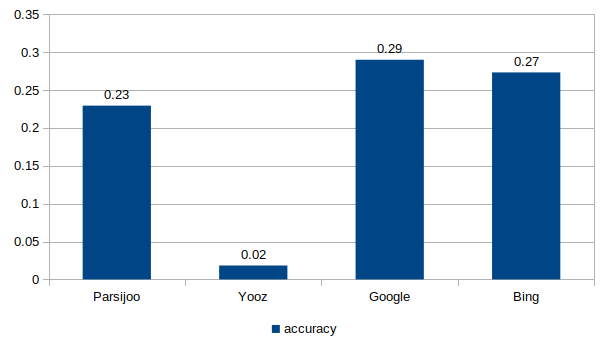} }}%
    \subfloat[known items]{{\includegraphics[width=6.2cm]{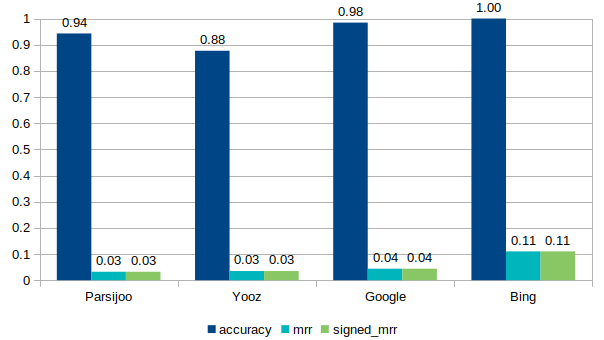} }}%
    \qquad
    \subfloat[semi informational]{{\includegraphics[width=6.2cm]{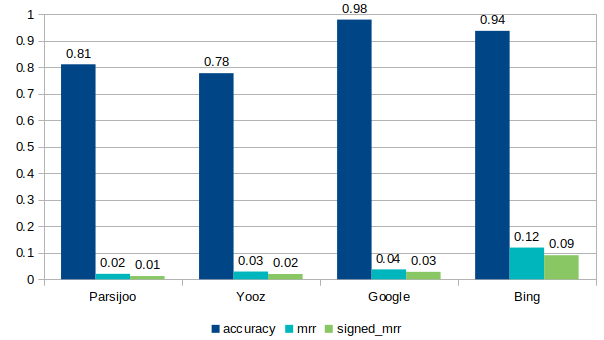}}}
    \subfloat[sum up]{{\includegraphics[width=6.2cm]{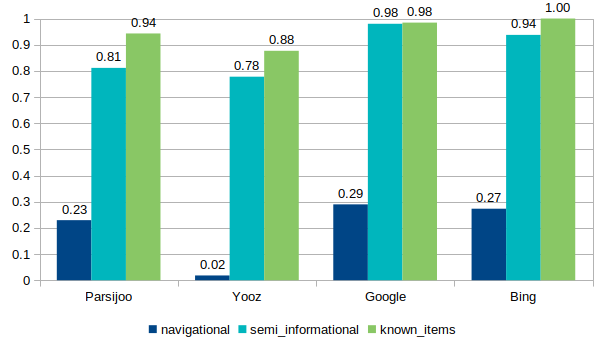}}}
    \caption{Rank evaluation}%
    \label{fig7}%
\end{figure}

is the main power of Google by which can attract much higher number of users than other search engines like Yooz and Parsijoo. Figure~\ref{fig7}:c states that Google and Bing are similar in terms of their accuracy score; however, a higher MRR and signed MRR score for Bing shows its results' higher quality in contrast with Google's. There is a similar relationship between Parsijoo and Yooz in both explained terms. Furthermore, both international engines outperformed two national ones in terms of all determining metrics. 

\subsubsection{Rank} the comparison of the overall score of the rank component of search engines states that Google has done a great job by making a great difference against other systems. National search engines can build a roadmap based on the current results to improve their systems to gain the ability to answer their users' needs. 

\section{Discussion and Future work}
As discussed throughout this paper, previous work has some weak points like small query set, narrow range of query types, subjective evaluation, high cost of evaluation and re-evaluation, and finally not evaluating all the components of a search engine. This work attempted to address all these problems with suitable, robust, low cost and reusable solutions. Results elicit the truth weak and power points of each search engine; not only according to each other but also to the true definition of each task that a search engine should address to provide appropriate answers to their users' queries. 

Designing an automatic system for building the query-set is one of the further work that can be investigated to improve this work. The structure of queries can be learned by text-mining methods, after training a robust model it can extract new queries using an unlabeled corpus. Moreover, some other evaluation metrics can also be used to analyze studying systems in a better level of detail. 

\section*{Acknowledgment}
We are really thankful for Dr. MohammadAli Abam's cooperation and support in managing this project. Truly, he made a great benefit for accomplishing this project. Additionally, This project was supported by Iran Telecommunication Research Center with project id 901952720. 

%
%
%
%


\end{document}